\documentclass[aps,twocolumn,amsmath,amssymb]{revtex4-2}
\usepackage[dvipdfmx]{graphicx} 
\usepackage{dcolumn}
\usepackage{bm}
\usepackage{braket}
\usepackage{multirow}
\usepackage[usenames]{color}
\usepackage{hyperref}
\usepackage{bbm}
\hypersetup{
colorlinks = true, 
urlcolor = blue, 
linkcolor = blue, 
citecolor = blue
}

\begin{document}

\title{
Fluctuation exchange study on the electron-hole asymmetry of the superconductivity across 1/3 filling in the trilayer Hubbard model
}

\author{
Yushi Yamada,$^1$ Masataka Kakoi,$^1$ Masayuki Ochi,$^{1,2}$ and Kazuhiko Kuroki$^1$
}

\affiliation{
$^1$Department of Physics, The University of Osaka, Toyonaka, Osaka 560-0043, Japan \\
$^2$Forefront Research Center, The University of Osaka, Toyonaka, Osaka 560-0043, Japan
}

\date{\today}

\begin{abstract}
We study within the fluctuation exchange approximation the trilayer Hubbard model where three layers of the Hubbard model are coupled by large interlayer hoppings so that the overlap of the bonding, nonbonding, and antibonding bands is relatively small. We pay special attention to the band fillings close to 1/3, for which the bonding and nonbonding bands as a whole are close to  half filling. For relatively small values of the onsite $U$, superconductivity roughly exhibits electron-hole symmetric behavior, as expected for a nearly half-filled two-band system. By contrast, an asymmetry appears when $U$ becomes large, where superconductivity is more favored in the hole-doped regime, i.e., in the regime where electrons are removed from 1/3 filling. We attribute this asymmetry to the asymmetric renormalization of the bonding, nonbonding, and antibonding bands when $U$ is large.
\end{abstract}

\maketitle


\section{Introduction}
Studying the possibility of high $T_{\rm c}$ superconductivity in purely theoretical models can provide insight into the search for actual high $T_{\rm c}$ materials. For example, a purely theoretical bilayer Hubbard model, where two layers of the Hubbard model are coupled by vertical interlayer hoppings, was shown to exhibit superconductivity with a higher $T_{\rm c}$ compared to a single band Hubbard model~\cite{Kuroki_2002,Maier_2011,Nakata_2017,Nomura_2026}. Such studies have led to a theoretical proposal for realizing the bilayer Hubbard model in a bilayer nickelate La$_3$Ni$_2$O$_7$~\cite{Nakata_2017}, which is now known to be a high $T_{\rm c}$ superconductor under pressure~\cite{H-Sun_2023,J-Hou_2023,G-Wang_2024,Y-Zhang_2024,N-Wang2024,F-Li_Sm96K_2026} or at ambient pressure for thin films~\cite{EK-Ko_2025,G-Zhou_2025, G-Zhou_films63K_2026}. This experimental discovery has sparked a huge wave of theoretical studies on the material~\cite{Z-Luo_2023,QG-Yang_2023,Sakakibara_2024a,Christiansson_2023,Lechermann_2023,Y-Shen_2023,YB-Liu_2023,C-Lu_2024,H-Oh_2023,XZ-Qu_2024,Kaneko_2024,Kakoi_2024,Kaneko_2025,Kamiyama_2025,Ushio_2026,Asai_arXiv,Watanabe_arXiv}. The discovery of superconductivity in the bilayer nickelate was followed by the discovery of superconductivity under pressure in the trilayer version of the nickelate, La$_4$Ni$_3$O$_{10}$~\cite{Sakakibara_2024b,Q-Li_2024,Y-Zhu_2024,Nagata_2024,M-Zhang_2025,E-Zhang_2025,D-Peng_2026}, which was theoretically predicted by a combination of first principles calculation and fluctuation exchange (FLEX)  study~\cite{Sakakibara_2024b}. This experimental discovery has also stimulated a large number of theoretical studies~\cite{Tian_2024,Qiong_2024,CQ-Chen_2024,J-Huang_2024,Ming-Zhang_2024,LaBollita_2024,Ming-Zhang_2025,C-Lu_2025,S-Zheng_2026,J-Chen_2026}.

Apart from the trilayer nickelate superconductor, studying a purely theoretical trilayer Hubbard model, where three layers of the Hubbard model are coupled by vertical interlayer hoppings $t_\perp$, may provide insight into not just the nickelate but other trilayer materials as well. In particular, if we focus on cases where the interlayer hopping is large so that the overlap between the bonding, nonbonding, and antibonding bands is small or even does not exist, the band filling near 1/3 would be of special interest because the bonding and nonbonding bands as a whole would be nearly half filled, where high $T_{\rm c}$ superconductivity is expected in the bilayer Hubbard model (with just bonding and antibonding bands)~\cite{Kuroki_2002,Maier_2011,Nakata_2017,Nomura_2026}.
In fact, a weak coupling theory for the one-dimensional version of the trilayer Hubbard model, namely, the three-leg Hubbard ladder, shows that the system is a spin-gapped insulator at exactly 1/3 filling when the interchain hopping is large enough~\cite{Arrigoni_1996a,*Arrigoni_1996b}. The weak coupling theory also shows that there is an ``electron-hole  symmetry'' across 1/3 filling, namely, cases where electrons are removed from or added to 1/3 filling have similar properties, similarly as in the two-leg Hubbard ladder across half filling. Strong coupling studies including our recent work on the three-leg $t$-$J$ ladder model with large interchain hopping also show that 1/3 is a special band filling, but indicate that there is actually an electron-hole asymmetry across 1/3 filling~\cite{Kagan_1999,Yamada_arXiv}. 

Given this background, here we study the trilayer Hubbard model, where the  interlayer hopping $t_{\perp}$ is taken to be four times larger than the nearest neighbor intralayer hopping. We adopt the combination of the FLEX approximation and the linearized Eliashberg equation to study superconductivity, where we pay special attention to the band fillings close to 1/3. For relatively small values of the on-site $U$, the tendency toward superconductivity roughly exhibits electron-hole symmetry across 1/3 filling, similar to the conclusion of the weak coupling theory for the three-leg Hubbard ladder. By contrast, an asymmetry appears when $U$ becomes large, where superconductivity is more favored in the hole-doped regime, i.e., in the regime where electrons are removed from 1/3 filling. Interestingly, this asymmetry across 1/3 filling in the tendency towards pairing resembles that obtained for a three-leg $t$-$J$ ladder in our recent density matrix renormalization group (DMRG) study~\cite{Yamada_arXiv}.


\section{Model and Method}

\begin{figure}[t]
    \centering
    \includegraphics[width=1.0\linewidth]{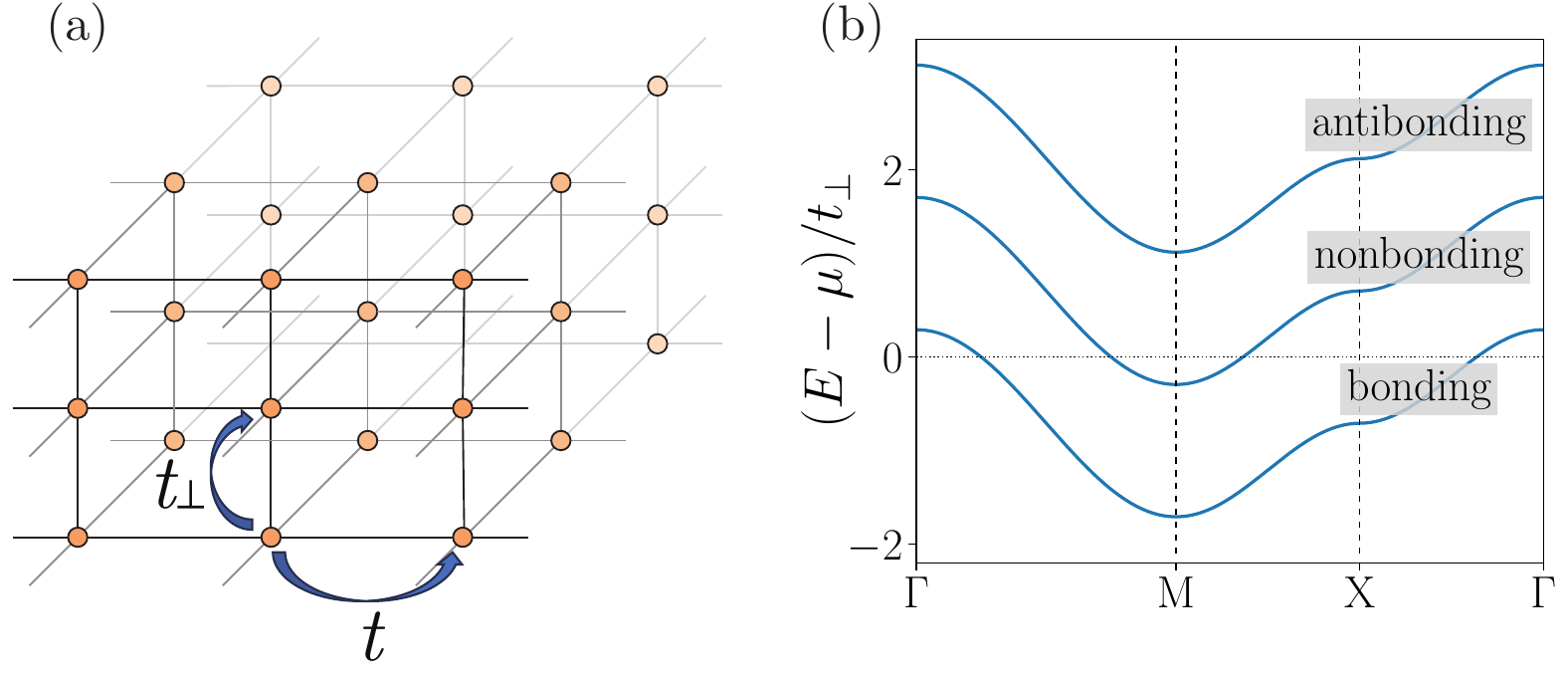}
    \caption{(a) Real-space structure of the trilayer square lattice. (b) Tight-binding band structure for $t/t_{\perp}=0.25$, with the chemical potential $\mu$ set to $1/3$-filling.}
    \label{fig:lattice_band}
\end{figure}

We study the Hubbard model of electrons on the trilayer square lattice [see Fig.~\ref{fig:lattice_band}(a)], whose Hamiltonian reads
\begin{align}\label{eq:trilayer_Hubbard_model}
  \hat{H} = & ~t\sum_{\langle\bm{r},\bm{r}'\rangle} \sum_{l} 
  \sum_{\sigma} \hat{c}^\dagger_{\bm{r}, l, \sigma} \hat{c}^{}_{\bm{r}', l, \sigma} \notag \\
& +t_\perp\sum_{\bm{r}}\sum_{\langle l,l' \rangle}\sum_{\sigma} \hat{c}^\dagger_{\bm{r}, l, \sigma} \hat{c}^{}_{\bm{r}, l', \sigma} \notag \\
& + U\sum_{\bm{r}}\sum_{l}\hat{n}_{\bm{r}, l, \uparrow}\hat{n}_{\bm{r}, l, \downarrow},
\end{align}
where $\hat{c}^\dagger_{\bm{r}, l, \sigma}\ (\hat{c}^{}_{\bm{r}, l, \sigma})$ is the creation (annihilation) operator of an electron with spin $\sigma$~($=\uparrow,\downarrow$) at site $\bm{r}$ on layer $l$~$(=1,2,3)$, and $\hat{n}_{\bm{r}, l, \sigma}=\hat{c}^\dagger_{\bm{r}, l, \sigma} \hat{c}^{}_{\bm{r}, l, \sigma}$ is the number operator. 
$\langle \bm r,\bm r' \rangle$ and $\langle l,l' \rangle$ denote nearest-neighbor pairs of sites and layers, respectively.
$t$ and $t_{\perp}$ denote the intralayer and interlayer hopping integrals, and $U$ is the onsite Coulomb repulsion.

The tight-binding part of Eq.~\eqref{eq:trilayer_Hubbard_model} ($t$ and $t_{\perp}$ terms) can be written as
\begin{equation}
    \hat{H}_0 = \sum_{\bm{k}}\sum_{\langle l,l' \rangle}\sum_\sigma [H_0(\bm{k})]^{}_{ll'}\hat{c}^\dagger_{\bm{k}, l, \sigma} \hat{c}^{}_{\bm{k}, l', \sigma},
\end{equation}
where $\bm{k} = (k_x, k_y)$ denotes a Bloch wave vector, and the matrix $H_0(\bm{k})$ is
\begin{equation}
    H_0(\bm{k}) = \left(
    \begin{array}{ccc}
        t\gamma_{\bm{k}} & t_{\perp}        & 0  \\
        t_{\perp}        & t\gamma_{\bm{k}} & t_{\perp}  \\
        0                & t_{\perp}        & t\gamma_{\bm{k}}
    \end{array}
    \right),
\end{equation}
with
\begin{equation}
    \gamma_{\bm{k}} = 2(\cos k_x + \cos k_y).
\end{equation}
Diagonalizing $H_0(\bm{k})$ yields eigenvalues
\begin{align}
    \varepsilon_{\bm{k},\mathrm{B}}  &= t\gamma_{\bm{k}} - \sqrt{2}t_{\perp}, \\
    \varepsilon_{\bm{k},\mathrm{NB}} &= t\gamma_{\bm{k}}, \\
    \varepsilon_{\bm{k},\mathrm{AB}} &= t\gamma_{\bm{k}} + \sqrt{2}t_{\perp},
\end{align}
corresponding to the bonding (B), nonbonding (NB), and antibonding (AB) bands, see Fig.~\ref{fig:lattice_band}(b) for the band structure. 
The corresponding eigenvectors are
\begin{equation}
\begin{gathered}
    \ket{\mathrm{B}} = \frac{1}{2} \left(
    \begin{array}{c}
         1  \\
         -\sqrt{2}  \\
         1
    \end{array}
    \right),\quad
    \ket{\mathrm{NB}} = \frac{1}{\sqrt{2}} \left(
    \begin{array}{c}
         1  \\
         0  \\
         -1
    \end{array}
    \right),\\
    \ket{\mathrm{AB}} = \frac{1}{2} \left(
    \begin{array}{c}
         1  \\
         \sqrt{2}  \\
         1
    \end{array}
    \right),
\end{gathered}
\end{equation}
respectively.

To investigate spin-fluctuation-mediated superconductivity, we adopt FLEX approximation~\cite{Bickers_1989,Dahm_1995} combined with the linearized Eliashberg equation. 
Using the self-energy calculated within FLEX and the resulting renormalized Green's function $G(\bm{k},i\omega_n)$, we solve the linearized Eliashberg equation
\begin{align}\label{eq:Eliashberg_eq}
    &\lambda \Delta_{ll'}(\bm{k},i\omega_n) \notag \\
    & = -\frac{T}{N}\sum_{\bm{k}'}\sum_{n'}\sum_{l_{1},l_{2},l_{3},l_{4}}\Gamma_{ll_{1},l_{4}l'}(\bm{k}\!-\!\bm{k}',i\omega_n\!-\!i\omega_{n'})    \notag \\
    &\qquad \times G_{l_1l_2}(\bm{k}',i\omega_{n'})\Delta_{l_2l_3}(\bm{k}',i\omega_{n'})G_{l_4l_3}(-\bm{k}',-i\omega_{n'})
\end{align}
to obtain the eigenvalue $\lambda$ and the anomalous self-energy (gap function) $\Delta$. 
Here, $T$, $N$, and $\Gamma$ denote the temperature, the number of cells, and the pairing interaction, respectively.
We regard $\lambda$ calculated at a fixed temperature as the quantity representing how high the $T_{\rm c}$ of the system is.
We use $T/t_{\perp}=0.01$, $2\times4096$ Matsubara frequencies, and a $48\times48$~$\bm{k}$ mesh throughout the paper.
For $\Gamma$, we use a spin-singlet pairing interaction
\begin{align}
    \Gamma(\bm{q},i\Omega_m)
    &=\frac32U^{\rm s}\chi^{\rm s}(\bm{q},i\Omega_m)U^{\rm s} 
    - \frac12U^{\rm c}\chi^{\rm c}(\bm{q},i\Omega_m)U^{\rm c} \notag\\
    &\quad+ \frac12\left(U^{\rm s}\!+\!U^{\rm c}\right),
\end{align}
where the spin and charge susceptibilities ($\chi^{\rm s}$ and $\chi^{\rm c}$) are
\begin{align}
    \label{eq:chi_S}
    \chi^{\rm s}(\bm{q},i\Omega_m) &= \chi^0(\bm{q},i\Omega_m)\left[
    \mathbbm{1}-U^{\rm s}\,\chi^0(\bm{q},i\Omega_m)
    \right]^{-1},\\[1pt]
    \chi^{\rm c}(\bm{q},i\Omega_m) &= \chi^0(\bm{q},i\Omega_m)\left[
    \mathbbm{1}+U^{\rm c}\,\chi^0(\bm{q},i\Omega_m)
    \right]^{-1},
\end{align}
with the irreducible susceptibility defined as
\begin{align}
    &\chi^0_{l_1l_2,l_3l_4}(\bm{q},i\Omega_m) \notag\\
    &= -\frac{T}{N} \sum_{\bm{k},n} 
    G_{l_1l_3}(\bm{k},i\omega_{n})
    G_{l_4l_2}(\bm{k}\!+\!\bm{q},i\omega_{n}\!+\!i\Omega_m).
\end{align}
Since we only consider the onsite Coulomb interaction,
$U^{\rm s}_{l_1l_2,l_3l_4}=U^{\rm c}_{l_1l_2,l_3l_4}
=U\delta_{l_1,l_2}\delta_{l_1,l_3}\delta_{l_1,l_4}$.


\section{Results}

We mainly focus on the case of $t/t_{\perp}=0.25$, the band structure for which is shown in Fig.~\ref{fig:lattice_band}(b). The $t/t_{\perp}$ dependence of our calculation results is discussed in Sec.~\ref{sec:tperp}.

\subsection{The eigenvalue $\lambda$ and the gap function $\Delta$}

\begin{figure}[t]
    \centering
    \includegraphics[width=1.0\linewidth]{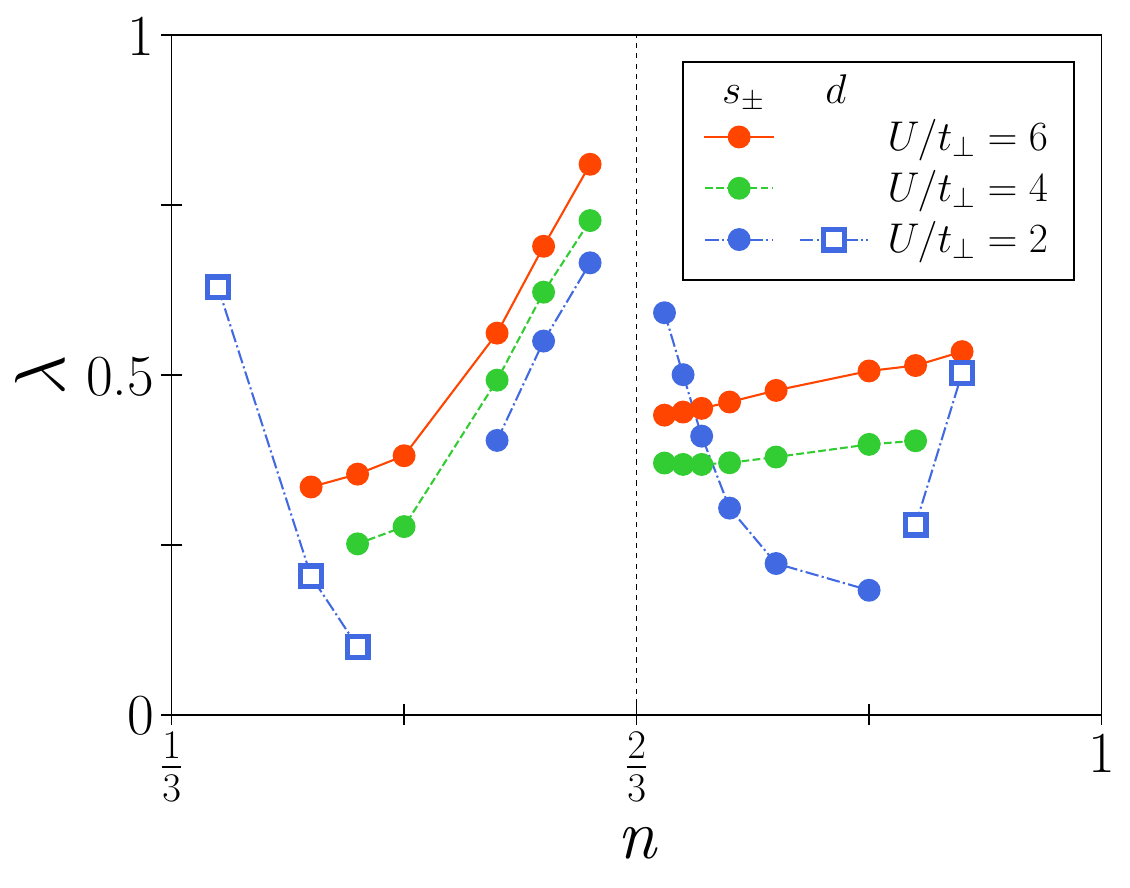}
    \caption{Band filling $n$ dependence of the eigenvalue $\lambda$ of the linearized Eliashberg equation~\eqref{eq:Eliashberg_eq} for various values of $U/t_{\perp}=2,4,6$ at $t/t_{\perp}=0.25$. 
    Band filling $n=2/3$ corresponds to $1/3$-filling.
    Circles denote points where the anomalous self-energy (gap function) $\Delta$ has $s$-wave symmetry, while open squares denote $d$-wave symmetry.}
    \label{fig:lambda_vs_n}
\end{figure}

Figure~\ref{fig:lambda_vs_n} shows the eigenvalue $\lambda$ obtained by solving the linearized Eliashberg equation~\eqref{eq:Eliashberg_eq} for various $U/t_{\perp}$ and the band filling $n$.
Here, the band filling is defined as the number of electrons per site, by which $n=2/3$ corresponds to $1/3$-filling.
Interestingly, while $\lambda$ is symmetric about $n=2/3$ for $U/t_{\perp}=2$, $\lambda$ in the electron-doped region is strongly suppressed for $U/t_{\perp}=4$ and $6$.
This situation is reminiscent of the electron--hole asymmetry found in the three-leg ladder with strong interchain coupling.
In the weak coupling limit, the three-leg ladder with strong interchain coupling hosts the C1S0 phase, which is favorable for superconductivity, both in electron- and hole-doped regions around the $1/3$-filling~\cite{Arrigoni_1996a,*Arrigoni_1996b,Kimura_1996,*Kimura_1998,H-Lin_1997}. On the other hand, in the strong coupling limit, it was shown that the pair correlation is strongly suppressed in the hole-doped region, resulting in the electron--hole asymmetry in superconductivity~\cite{Kagan_1999,Yamada_arXiv}.
These observations in the three-leg ladder are consistent with our calculation results in the trilayer square lattice with strong interlayer hopping.
It is noteworthy that FLEX calculations, despite being categorized as a weak-coupling theory, can partially capture superconductivity in the strong interaction regime, as demonstrated in, e.g., bilayer systems.
Namely, the gap function obtained for the bilayer Hubbard model using FLEX exhibits almost no momentum dependence~\cite{Kuroki_2002}, indicating that the pairing, in real space, occurs basically in the interlayer intra-unit-cell channel, consistent with a real space picture in the strong coupling limit.

\begin{figure*}[t]
    \centering
    \includegraphics[width=1.0\linewidth]{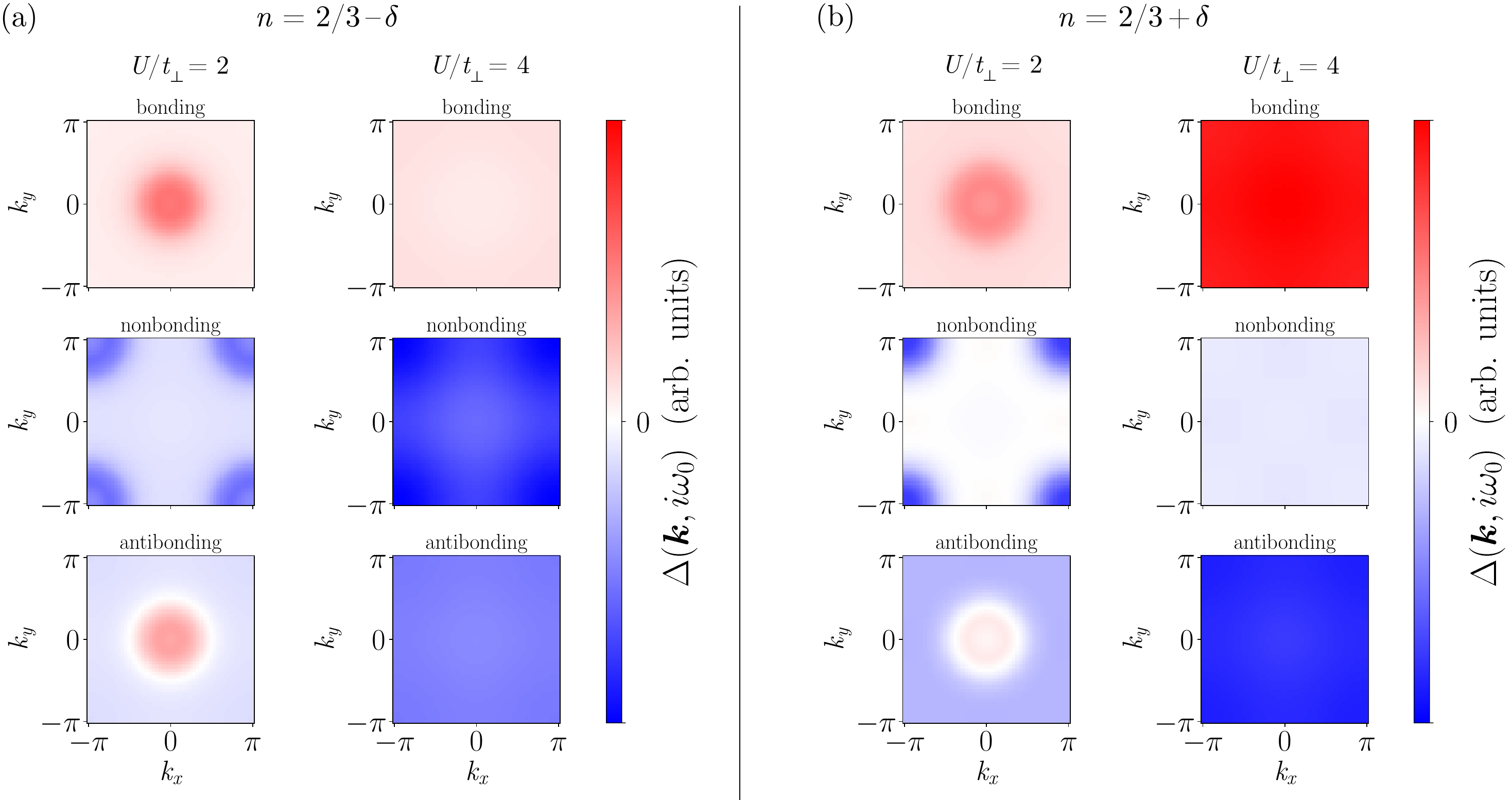}
    \caption{Band representations of the gap function $\Delta(\bm{k},i\omega_0)$ at the lowest Matsubara frequency $i\omega_0=i\pi k_{\rm B}T$, obtained for $U/t_{\perp}=2,4$ at (a)~$n=2/3 - \delta$ and (b)~$n=2/3 + \delta$ with $\delta = 0.0333$.}
    \label{fig:gapfunction_band}
\end{figure*}

Next, we discuss the gap functions.
As shown in Fig.~\ref{fig:lambda_vs_n}, the gap functions around 1/3-filling are mostly $s$-wave, while $d$-wave solutions are found around $n=1/3$ and $1$ for a small $U/t_{\perp}$, which can be understood as they correspond to half-filling of the bonding and nonbonding bands, respectively.
Figure~\ref{fig:gapfunction_band} shows the gap functions $\Delta(\bm{k},i\omega_n)$ at the lowest Matsubara frequency $i\omega_0=i\pi k_{\rm B}T$ with band representation for the hole-doped ($n=2/3 - \delta$) and electron-doped ($n=2/3 + \delta$) cases with $\delta = 0.0333$.
As shown in Fig.~\ref{fig:gapfunction_band}(a), for the hole-doped case with $U/t_{\perp}=2$, the gap function exhibits a large weight near the $\Gamma$ point for the bonding band and the $\mathrm{M}$ point for the nonbonding band, with opposite signs; we refer to this structure as an $s_\pm$-wave.
This gap function suggests that low-energy pair scatterings between the bonding and nonbonding bands near the Fermi level play an important role in the enhancement of superconductivity.
In addition, relatively large positive values of the gap function near the $\Gamma$ point for the antibonding band indicate that interband scatterings between the nonbonding and antibonding bands may also contribute to superconductivity.
In contrast, for $U/t_{\perp}=4$, the gap function shows almost no ${\bm k}$ dependence, reflecting local pair formation in real space. The sign of the gap function suggests that pair scatterings between the bonding and nonbonding bands and/or those between the bonding and antibonding bands are important.
The gap functions for the electron-doped case are presented in Fig.~\ref{fig:gapfunction_band}(b).
For $U/t_{\perp}=2$, the gap function is very similar to that of the hole-doped case shown in Fig.~\ref{fig:gapfunction_band}(a), which is consistent with electron--hole symmetric behavior of $\lambda$.
On the other hand, for $U/t_{\perp}=4$, the large amplitude of the gap function in the bonding and antibonding bands suggests that pair scatterings between these two bands are important. 

\subsection{Spectral functions}

\begin{figure*}[tbp]
    \centering
    \includegraphics[width=\linewidth]{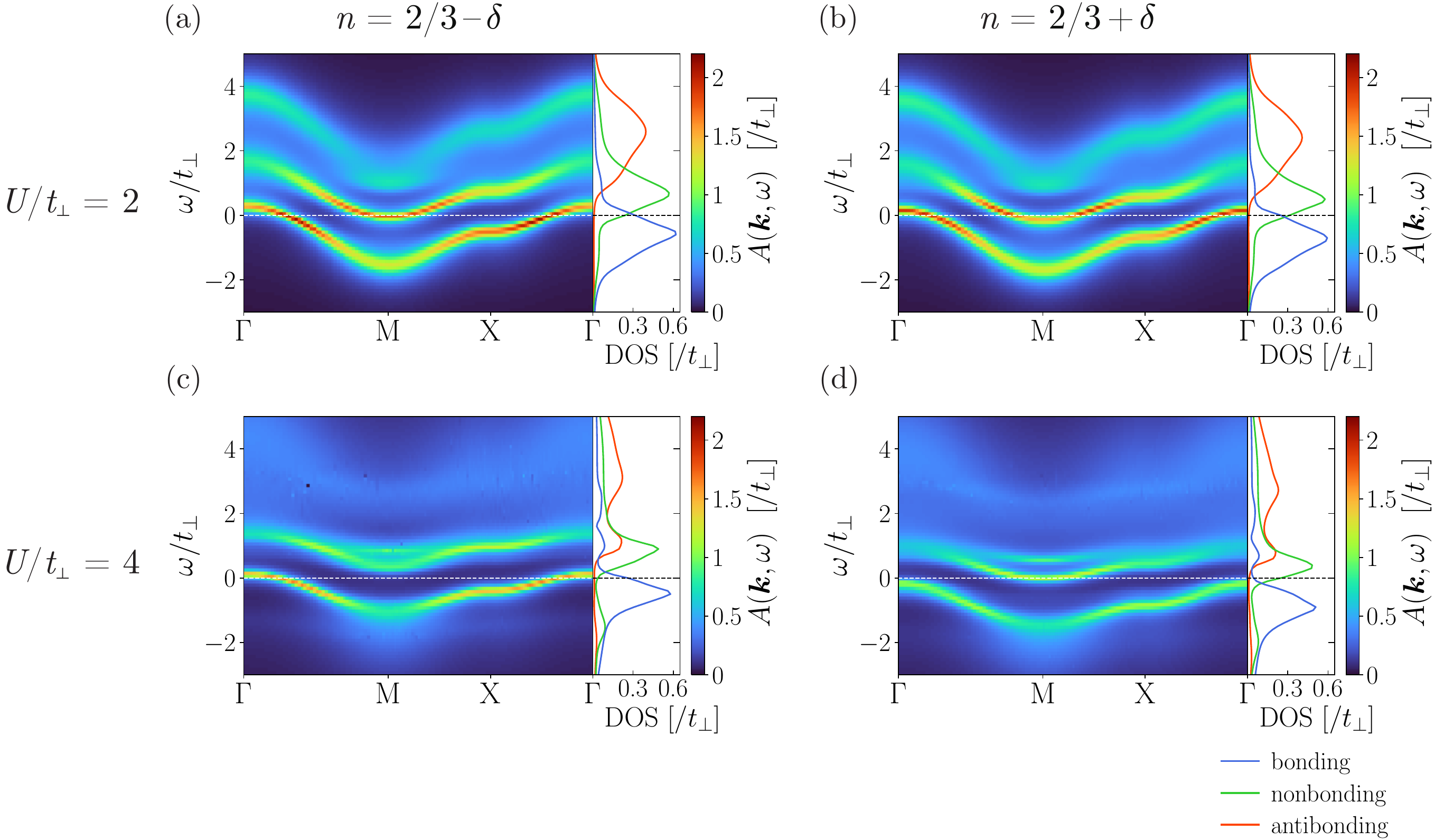}
    \caption{Spectral functions and partial density of states (DOS) for (a)~$n=2/3 - \delta$, $U/t_{\perp}=2$, (b)~$n=2/3 + \delta$, $U/t_{\perp}=2$, (c)~$n=2/3 - \delta$, $U/t_{\perp}=4$, and (d)~$n=2/3 + \delta$, $U/t_{\perp}=4$,
    with $\delta = 0.0333$.
    The blue, green, and red lines represent the partial DOS components of the bonding, nonbonding, and antibonding bands, respectively.}
    \label{fig:spectrum_DOS}
\end{figure*}

To elucidate the electron--hole asymmetry in $\lambda$ and $\Delta$, we calculate the spectral function 
\begin{equation}
A(\bm{k},\omega)=-\frac{1}{\pi}\mathrm{Tr}\left[ \mathrm{Im}~G(\bm{k},\omega) \right],
\end{equation}
and the partial density of states (DOS), defined as
\begin{equation}
\frac{1}{N_{\bm{k}}}\sum_{\bm{k}}A_{\alpha\alpha}(\bm{k},\omega),
\end{equation}
as shown in Fig.~\ref{fig:spectrum_DOS}.
The spectral function is obtained via analytic continuation of the Green's function using Pad\'{e} approximation.

We first present the spectral functions and partial DOS for $U/t_{\perp}=2$ in Figs.~\ref{fig:spectrum_DOS}(a) and (b).
For both the (a) hole-doped and (b) electron-doped cases, the spectral function is similar to the non-interacting bands shown in Fig.~\ref{fig:lattice_band}, and the Fermi level intersects the bonding and nonbonding bands.
Thus, we can infer that Fermi surface nesting contributes to superconductivity, which provides a natural explanation of the electron--hole symmetric behavior of $\lambda$ and $\Delta$ in the weak $U$ regime.

We next present the spectral functions and partial DOS for $U/t_{\perp}=4$ in Figs.~\ref{fig:spectrum_DOS}(c) and (d).
Here, strong correlation effects substantially renormalize the band structure. As a result, an electron (hole) pocket does not form for hole- (electron-)doped regime.
In addition, the antibonding band approaches the Fermi level and partially overlaps with the nonbonding band, making the spectral functions resemble a bilayer band structure with a doubly degenerate upper band.
These features provide a natural explanation for the pronounced electron--hole asymmetry in $\lambda$ and $\Delta$.

It is instructive to compare a nearly $1/3$-filled trilayer with a nearly half-filled bilayer.
In the bilayer square lattice with strong interlayer spin fluctuations, $s_{\pm}$-wave superconductivity, in which the gap function changes sign between the bonding and antibonding bands, is strongly enhanced~\cite{Kuroki_2002,Maier_2011}.
When $U$ is comparable to the bandwidth, the gap function exhibits almost no ${\bm k}$ dependence~\cite{Kuroki_2002}, indicating local pair formation between the layers.
Such pairing is driven by finite-energy spin fluctuations~\cite{Kuroki_2005,Mishra_2016,Nakata_2017,Maier_2019,Matsumoto_2020,Kato_2020}, which relies on interband scatterings involving an incipient band rather than Fermi-surface nesting between electron and hole pockets.
This mechanism can work also in three-band systems~\cite{Kobayashi_2016,Aida_2024,Kakoi_2026}.
In our calculations for the nearly $1/3$-filled trilayer, both the nonbonding and antibonding bands can act as incipient bands in the hole-doped regime, as indicated in Fig.~\ref{fig:spectrum_DOS}(c), which should be favorable for superconductivity. 
On the other hand, in the electron-doped regime, as shown in Fig.~\ref{fig:spectrum_DOS}(d), the Fermi level lies within  the nonbonding band, whose renormalized band width is narrower than that of the bonding band. This situation is reminiscent of cases where bands with wide and narrow bare widths coexist, for which superconductivity is enhanced (suppressed) when the Fermi level lies within the wide (narrow) band~\cite{Kuroki_2005,Matsumoto_2020,Kato_2020,Aida_2024}.

\subsection{Real space picture of the Cooper pairs}

\begin{figure*}[t]
    \centering
    \includegraphics[width=1.0\linewidth]{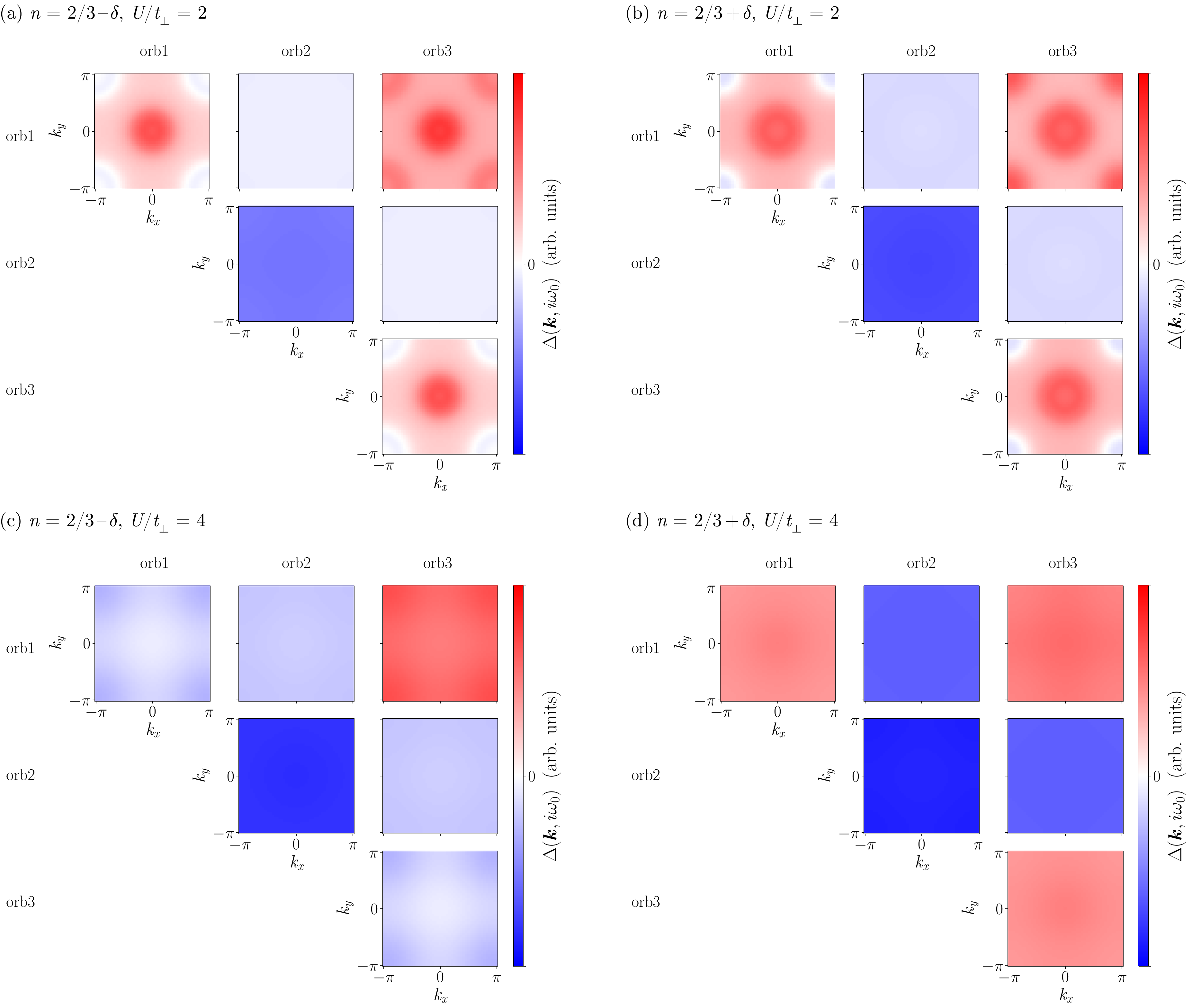}
    \caption{Orbital representations of the gap function for (a)~$n=2/3 - \delta$, $U/t_{\perp}=2$, (b)~$n=2/3 + \delta$, $U/t_{\perp}=2$, (c)~$n=2/3 - \delta$, $U/t_{\perp}=4$, and (d)~$n=2/3 + \delta$, $U/t_{\perp}=4$, with $\delta = 0.0333$.
    Orbitals $1$ and $3$ correspond to the outer sites, and orbital $2$ corresponds to the inner site.}
    \label{fig:gapfunction_orb}
\end{figure*}

To gain further insight, we present the gap functions in the orbital representation in Fig.~\ref{fig:gapfunction_orb}.
For the case of $U/t_{\perp}=2$, as shown in Figs.~\ref{fig:gapfunction_orb}(a) and (b), the large amplitudes of $\Delta_{13}$ and $\Delta_{11}(=\Delta_{33})$ correspond to interlayer pairing between outer layers and intralayer pairing within the outer layers, respectively.
These components exhibit some ${\bm k}$ dependence, indicating that the Cooper pairs extend in the in-plane directions. In contrast, the nearly ${\bm k}$ independent behavior of $\Delta_{22}$ means onsite pairing in the inner layer, as discussed later in this section.

For the case of $U/t_{\perp}=4$, as shown in Figs.~\ref{fig:gapfunction_orb}(c) and (d), all components are nearly ${\bm k}$ independent, indicating that the Cooper pairs are formed within the unit cell.
The large amplitudes of $\Delta_{13}$ and $\Delta_{22}$ correspond to interlayer pairing between the outer layers and intralayer pairing within the inner layer, respectively.
As shown in Fig.~\ref{fig:gapfunction_orb}(d), in the electron-doped regime, $\Delta_{12}$ and $\Delta_{23}$ also exhibit large amplitudes, which is consistent with sizable interband scatterings between the bonding and antibonding bands as discussed in connection with Fig.~\ref{fig:gapfunction_band}.

\begin{figure}[tbp]
    \centering
    \includegraphics[width=0.75\linewidth]{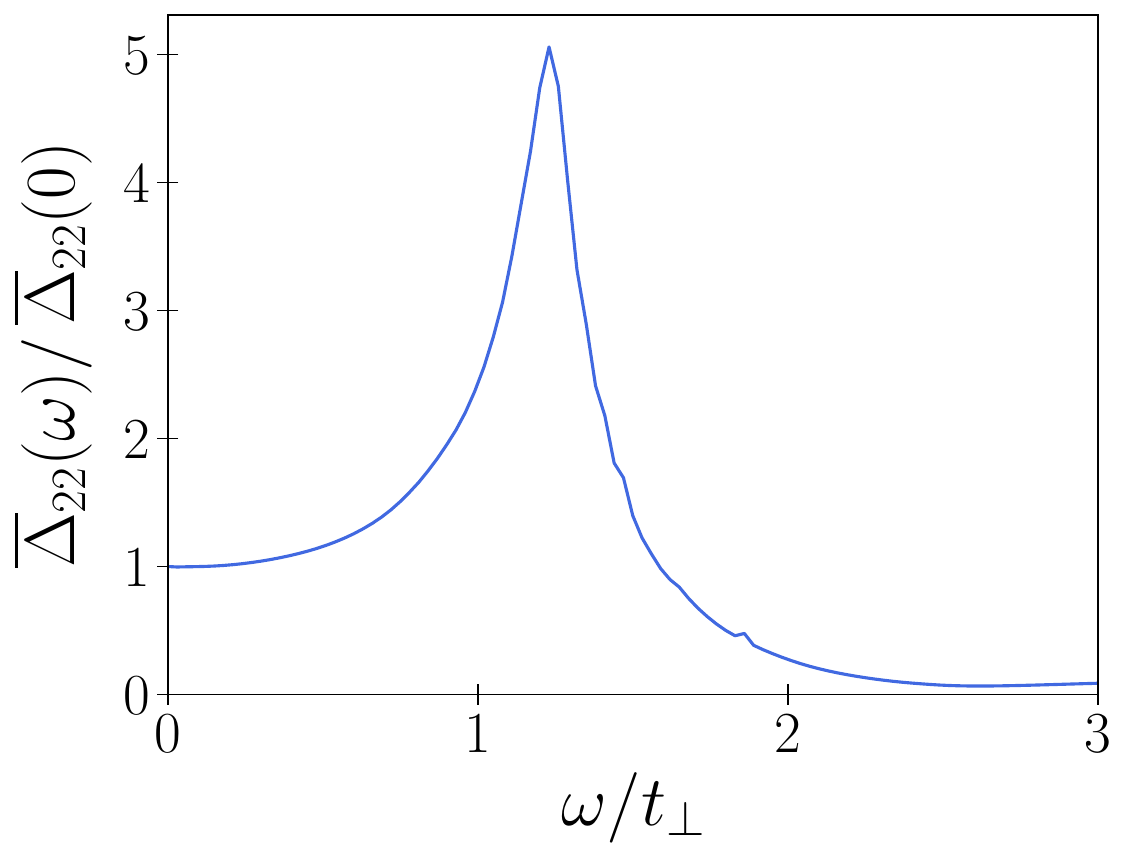}
    \caption{$\omega$ dependence of the $\bm{k}$-integrated absolute value of the gap function ($22$ component), defined as $\overline{\Delta}_{22}(\omega) := \sum_{\bm{k}}|\Delta_{22}(\bm{k},\omega)|$, obtained for $n=2/3 - \delta$ ($\delta = 0.0333$) and $U/t_{\perp}=4$.}
    \label{fig:gap_function_pade}
\end{figure}

In Fig.~\ref{fig:gapfunction_orb}, the ${\bm k}$-independent behavior of $\Delta_{22}$ indicates onsite pairing, which would appear to be unfavorable for superconductivity since Cooper pairing is expected to be suppressed by the onsite repulsion $U$.
This apparent contradiction can be understood from the frequency dependence of the gap function.
Fig.~\ref{fig:gap_function_pade} presents the ${\bm k}$-integrated gap function,
\begin{equation}
 \overline{\Delta}_{22}(\omega) := \sum_{\bm{k}}|\Delta_{22}(\bm{k},\omega)|,
\end{equation}
plotted as a function of real frequency $\omega$ for $n=2/3 - \delta$ ($\delta = 0.0333$) and $U/t_{\perp}=4$.
This quantity is obtained via analytic continuation and normalized by $\overline{\Delta}_{22}(0)$.
This function exhibits a peak around $\omega/t_{\perp} \sim 1.2$, suggesting retarded pairing, whereby the Cooper pairs effectively avoid the strong onsite repulsion $U$.

\subsection{Electron--hole asymmetry in low-energy spin fluctuations}

The electron--hole asymmetry around $1/3$-filling is also reflected in the spin fluctuations.
The spin susceptibilities presented in this section are obtained via analytic continuation, as in the previous sections.

\begin{figure}[!t]
    \centering
    \includegraphics[width=\linewidth]{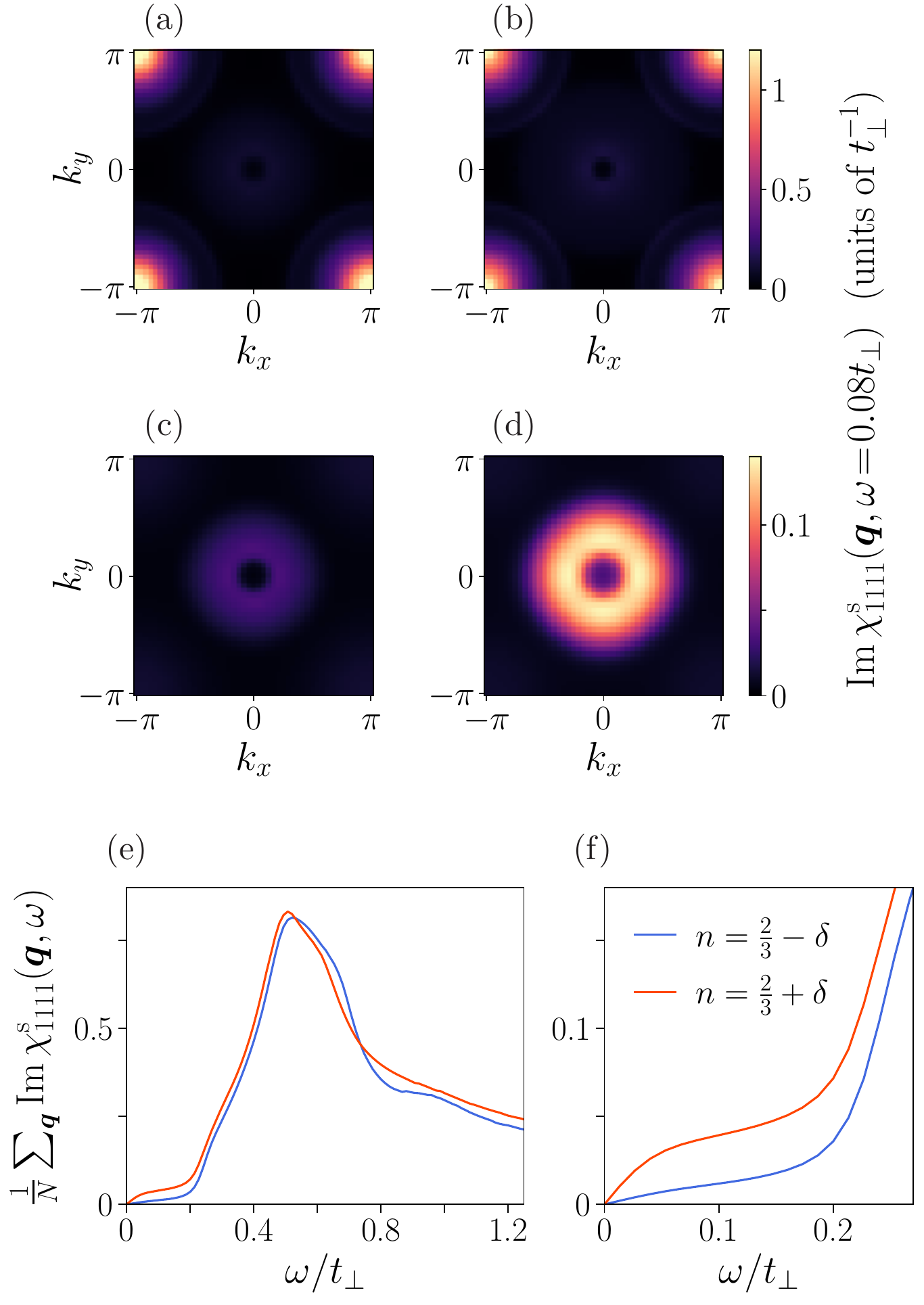}
    \caption{(a)--(d)Imaginary part of the $1111$ component of the dynamical spin susceptibility $\chi^{\rm s}_{1111}(\bm{q},\omega)$ at fixed $\omega=0.08t_{\perp}$ for (a)~$n=2/3 - \delta$, $U/t_{\perp}=2$, (b)~$n=2/3 + \delta$, $U/t_{\perp}=2$, (c)~$n=2/3 - \delta$, $U/t_{\perp}=4$, and (d)~$n=2/3 + \delta$, $U/t_{\perp}=4$, with $\delta = 0.0333$.
    (e) $\Omega$ dependence of the ${\bm q}$-integrated $\mathrm{Im}\,\chi^{\rm s}_{1111}(\bm{q},\omega)$ for $U/t_{\perp}=4$, corresponding to panels (c) and (d).
    (f) An enlarged view of (e).}
    \label{fig:chiS}
\end{figure}

Figures~\ref{fig:chiS}(a)--(d) show the imaginary part of the dynamic spin susceptibility $\chi^{\rm s}(\bm{q},\omega)$ [see, Eq.~\eqref{eq:chi_S}], at a representative low energy $\omega$.
Here, we focus on $\chi^{\rm s}_{1111}$ corresponding to the outer-layer component.
For $U/t_{\perp}=2$, as shown in Figs.~\ref{fig:chiS}(a) and \ref{fig:chiS}(b), $\mathrm{Im}\,\chi^{\rm s}$ reaches its maximum near $\bm{q}=(\pi,\pi)$ in both the hole- and electron-doped regimes. This behavior suggests that low-energy spin fluctuations associated with scattering between electron and hole pockets enhance superconductivity.
In contrast, for $U/t_{\perp}=4$, as shown in Figs.~\ref{fig:chiS}(c) and \ref{fig:chiS}(d), $\mathrm{Im}\,\chi^{\rm s}$ is an order of magnitude smaller than that for $U/t_{\perp}=2$, indicating a strong suppression of low-energy spin fluctuations.
Still, $\mathrm{Im}\,\chi^{\rm s}$ is larger in the electron-doped regime than in the hole-doped regime, around the $\Gamma$ point.
To examine this in more detail, we present the frequency dependence of the ${\bm q}$-integrated spin susceptibility in Figs.~\ref{fig:chiS}(e) and (f).
As shown in Fig.~\ref{fig:chiS}(f), the low-energy spin fluctuations for $\omega/t_{\perp}\lesssim0.2$ are more pronounced in the electron-doped regime.

The relationship between the electron--hole asymmetry in low-energy spin fluctuations and that in $\lambda$ can be understood as follows.
It has been pointed out that low-energy spin fluctuations can sometimes cause pair breaking~\cite{Millis_1988,Matsumoto_2020,Kato_2020,Kouchi_2022}.
Specifically, the contributions from pair-breaking and pairing-effective spin fluctuations scale as $1/\omega^4$ and $1/\omega^2$, respectively. 
Thus, the strong low-energy spin fluctuations in the electron-doped regime can result in the reduction of $\lambda$.

The enhanced low-energy spin fluctuations in the electron-doped regime can also be understood from a real-space picture.
When $t_{\perp}/t$ is large, a spin singlet ($S=0$) state for the total spin in the unit cell are formed at $1/3$-filling, where two electrons occupy each rung (interlayer pair).
In the hole-doped regime, two doped holes tend to occupy the same rung, allowing the system to maintain $S=0$ configurations on the remaining rungs.
In contrast, in the electron-doped regime, it is energetically more favorable to form an $S=1/2$ state rather than placing two additional (four in total) electrons on the same rung and forming an $S=0$ state, in order to avoid the strong onsite repulsion $U$.
As a result, the spin gap is more easily suppressed in the electron-doped regime than in the hole-doped regime, leading to more pronounced low-energy spin fluctuations in the former.
This picture is supported by the DMRG study of the three-leg ladder model~\cite{Yamada_arXiv}.

\subsection{$t/t_{\perp}$ dependence\label{sec:tperp}}

\begin{figure}[h]
    \centering
    \includegraphics[width=1.0\linewidth]{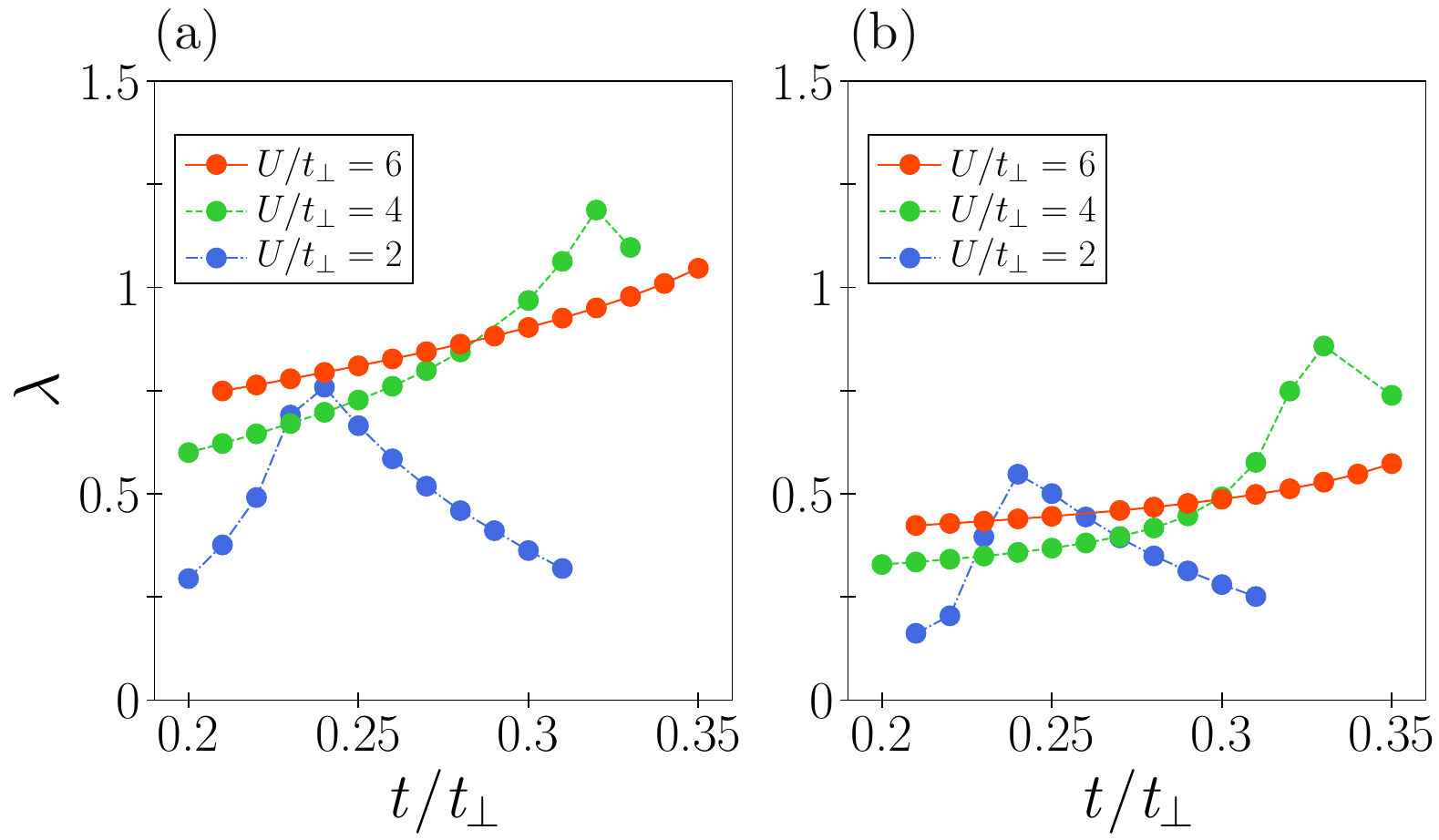}
    \caption{$t_{\perp}$ dependence of the eigenvalue $\lambda$ of the linearized Eliashberg equation~\eqref{eq:Eliashberg_eq} for various values of $U/t_{\perp}=2,4,6$ and for band fillings (a)~$n=2/3 - \delta$ and (b)~$n=2/3 + \delta$, with $\delta = 0.0333$.}
    \label{fig:lambda_vs_t}
\end{figure}

Figure~\ref{fig:lambda_vs_t} shows the $t/t_{\perp}$ dependence of the eigenvalue $\lambda$ calculated for band fillings $n=2/3 \pm \delta$ with $\delta = 0.0333$.
Overall, $\lambda$ is suppressed in the electron-doped regime, consistent with the electron--hole asymmetry observed at $t/t_{\perp}=0.25$ in the previous sections.
We also find that the value of $t/t_{\perp}$ at which $\lambda$ reaches its maximum increases with increasing $U/t_{\perp}$. This behavior can be understood as follows.
For $U/t_{\perp}=2$, values of $t/t_{\perp}$ slightly smaller than $0.25$ correspond to a situation in which the nonbonding (bonding) band can be regarded as an incipient band in the hole- (electron-)doped regime, as shown in Figs.~\ref{fig:spectrum_DOS}(a) and (b).
For larger $U/t_{\perp}$, the reduction of the bandwidth due to renormalization requires a larger $t/t_{\perp}$ for the system to satisfy the incipient-band condition, as inferred from Figs.~\ref{fig:spectrum_DOS} (c) and (d).
A similar behavior---namely, that the incipient-band condition determines the optimal hopping ratio for superconductivity---has also been reported in the bilayer Hubbard model~\cite{Matsumoto_2020}.


\section{Summary}

We have investigated spin-fluctuation-mediated superconductivity near $1/3$-filling in the trilayer Hubbard model with a large interlayer hopping $t_{\perp}$.
For small $U$, the eigenvalue of the linearized Eliashberg equation, $\lambda$, increases toward $1/3$-filling in both the hole- and electron-doped regimes.
In this case, the gap function has an $s_{\pm}$-wave symmetry, characterized by a sign change between different bands, together with some ${\bm k}$ dependence reflecting Fermi surface nesting.
In contrast, for large $U$, $\lambda$ increases toward $1/3$-filling in the hole-doped regime but is strongly suppressed in the electron-doped regime. 
In that case, the gap functions are nearly ${\bm k}$ independent, indicating local pair formation within the unit cell.
Our results are consistent with both weak-coupling~\cite{Arrigoni_1996a,*Arrigoni_1996b,Kimura_1996,*Kimura_1998,H-Lin_1997} and strong-coupling~\cite{Kagan_1999,Yamada_arXiv} approaches for the three-leg ladder model.
Superconductivity near $1/3$-filling in trilayer systems can also be understood by comparison with that in the half-filled bilayer model. In both systems, finite-energy spin fluctuations play a crucial role in enhancing $T_{\rm c}$ particularly in the strong-$U$ regime.

Considering that the stoichiometric trilayer nickelate (La,Pr)$_4$Ni$_3$O$_{10}$ can be regarded as a slightly hole-doped trilayer system relative to $1/3$-filling due to orbital hybridization~\cite{Sakakibara_2024b}, our finding that superconductivity is optimized in a slightly hole-doped trilayer relative to $1/3$-filling is particularly suggestive.
Further insight into multiorbital effects would be beneficial for a deeper understanding of superconductivity in trilayer nickelates.


\begin{acknowledgments}
    This work was supported by Grants-in-Aid for Scientific Research from JSPS, KAKENHI Grant No.~JP24K01333, JP25K08457, JP25K00959, JP26H02014, JP26K08179, and JST K Program Grant No. JPMJKP25Z3.
    M.K. thanks the Research Fellowship for Young Scientists (Grant No.~JP25KJ1758) for support.
\end{acknowledgments}





\bibliography{reference}

\end{document}